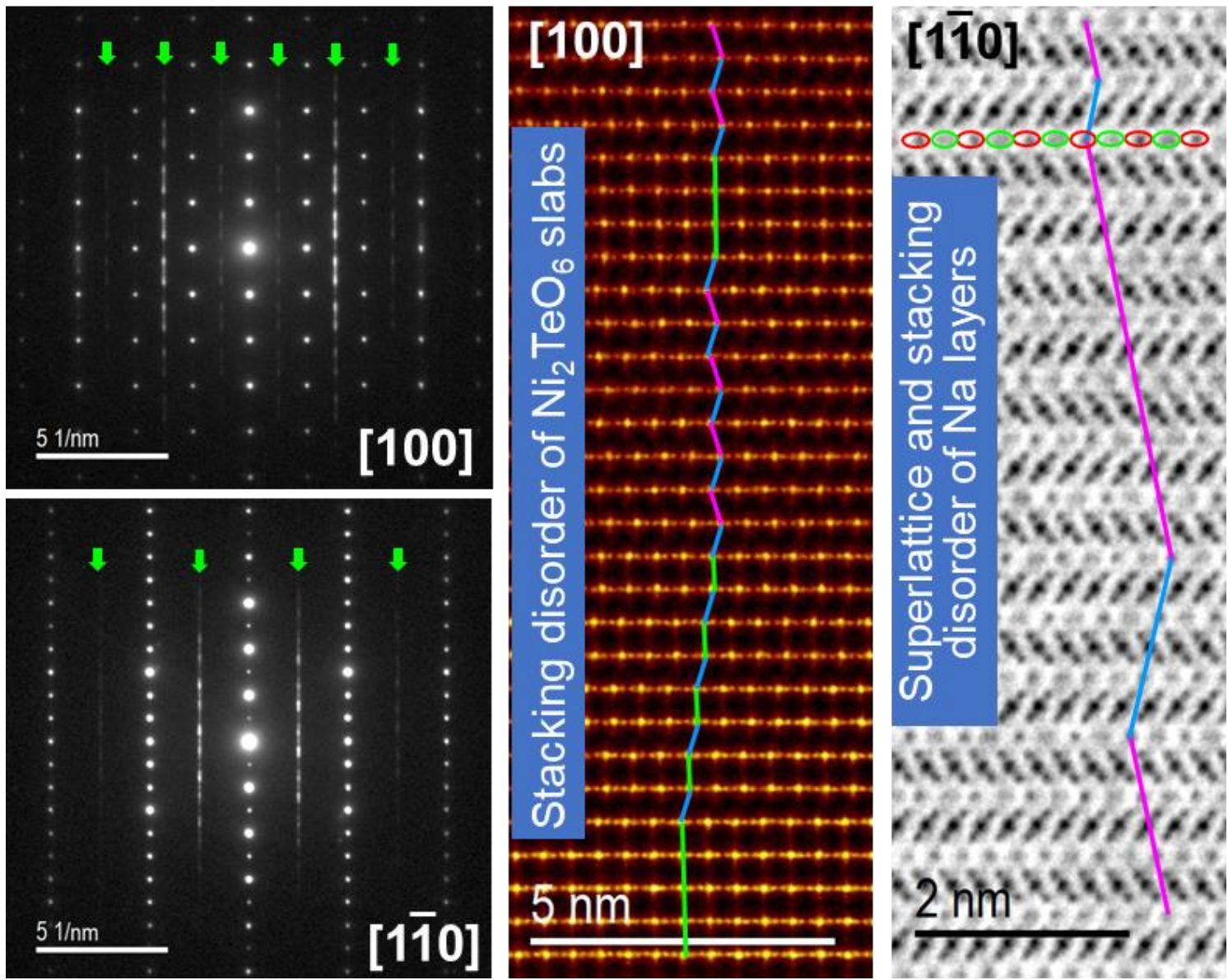

# Unveiling Structural Disorders in Honeycomb Layered Oxide: Na$_2$Ni$_2$TeO$_6$

Titus Masese[a,b], Yoshinobu Miyazaki[c], Josef Rizell[a,d], Godwill Mbiti Kanyolo[e], Teruo Takahashi[c], Miyu Ito[c], Hiroshi Senoh[a] & Tomohiro Saito[c]

[a] Research Institute of Electrochemical Energy, National Institute of Advanced Industrial Science and Technology (AIST), 1–8–31 Midorigaoka, Ikeda, Osaka 563–8577, JAPAN

[b] AIST–Kyoto University Chemical Energy Materials Open Innovation Laboratory (ChEM–OIL), Sakyo–ku, Kyoto 606–8501, JAPAN

[c] Tsukuba Laboratory, Technical Solution Headquarters, Sumika Chemical Analysis Service

(SCAS), Ltd.,Tsukuba, Ibaraki 300–3266, JAPAN

[d] Department of Physics, Chalmers University of Technology, SE–412 96 Göteborg, SWEDEN

[e] Department of Engineering Science, The University of Electro-Communications, 1–5–1 Chofugaoka, Chofu, Tokyo 182–8585, JAPAN

[*]Correspondence should be addressed to: Titus Masese

E-mail address:   titus.masese@aist.go.jp

Phone: +81–72–751–9224

Fax: +81–72–751–9609

**Abstract**

**Honeycomb layered oxides have garnered tremendous research interest in a wide swath of disciplines owing not only to the myriad physicochemical properties they exhibit, but also their rich crystal structural versatility. Herein, a comprehensive crystallographic study of a sodium-based $Na_2Ni_2TeO_6$ honeycomb layered oxide has been performed using atomic-resolution transmission electron microscopy, elucidating a plethora of atomic arrangement (stacking) disorders in the pristine material. Stacking disorders in the arrangement of honeycomb metal slab layers (stacking faults) occur predominantly perpendicular to the slabs with long-range coherence length and enlisting edge dislocations in some domains. Moreover, the periodic arrangement of the distribution of alkali atoms is altered by the occurrence of stacking faults. The multitude of disorders innate in $Na_2Ni_2TeO_6$ envisage broad implications in the functionalities of related honeycomb layered oxide materials and hold promise in bolstering renewed interest in their material science.**

**Keywords:** Transmission electron microscopy (TEM), Scanning transmission electron microscopy (STEM), Honeycomb layered oxides, Electron diffraction pattern, Atomic structure

# 1. <u>INTRODUCTION</u>

Recent years have seen a proliferation in the discovery and development of high-performance materials with exceptional physical, chemical and magnetic features, in an effort to satisfy the demands of ever-evolving cutting-edge technologies. It is in this vein that honeycomb layered oxides composed of coinage- or alkali metal atoms interspersed between layers of transition or heavy metal atoms arranged in a hexagonal pattern, have drawn momentous interest across multiple fields such as electrochemistry, material science, condensed matter physics *et cetera*. [1] Owing to their unique honeycomb configuration, a manifold of topological properties has emerged, demonstrating a cynosure of attributes such as rapid ionic conduction, fascinating electromagnetic and quantum capabilities, and diverse crystal chemistry amongst others. [1-18] Besides finding niche functionality in high voltage energy storage systems, [19–27] exotic magnetic behaviour, as envisaged by the Kitaev model, [28] makes them exemplar pedagogical platforms suited to the exploration of next-generation superconductors to topological quantum devices. [1]

A majority of the aforementioned nanostructured materials adopt chemical compositions such as $A_2M_2DO_6$, $A_3M_2DO_6$ or $A_4MDO_6$ wherein *A* represents an alkali- or coinage-metal species suchlike Li, Na, K, Cu, Ag *et cetera*., whereas *M* is a transition metal species such as Ni, Co, Zn, Co, *etc.* and *D* depicts a chalcogen or pnictogen metal species such as Te, Sb, Bi and so forth. [1,2,6,8,10-12,15,18-21,29-31] Due to the difference in the valency state and ionic radius of *M* and *D,* a distinct slab comprising $DO_6$ octahedra covalently-bonded with multiple $MO_6$ octahedra in a honeycomb configuration is formed. The oxygen atoms from these slabs in turn form weak coordinations with $A^+$ cations resulting into a lamellar framework of *A* alkali atoms sandwiched between parallel $MO_6$ and $DO_6$ octahedra slabs. [32]

Given that the interlayer distance is inversely proportional to the strength of the interlayer bonds, *A* atoms with a large Shannon-Prewitt radius typically form crystalline structures with weak interlayer bonds and *vice versa*, resulting in a diverse range of structural formations. [1] For instance, the smaller atomic radii Li cations in $Li_2Ni_2TeO_6$ result in stronger coordinations between Li atoms and oxygen atoms forming a tetrahedral structure with 2 repetitive honeycomb layers per unit cell, [20] whilst the larger atomic radii Na atoms in $Na_3Ni_2SbO_6$ and $Na_3Ni_2BiO_6$ result in weaker coordinations characterised by an octahedral structure with 3 repetitive honeycomb layers per unit cell. [1,22,24,25,27,33,34] Additionally, Na atoms in $Na_2Ni_2TeO_6$ exhibit much weaker coordinations with oxygen atoms resulting in a prismatic structure with 2 repetitive honeycomb layers. [6,10,16,] As a general rule of classification for the various arrangements (stacking) of atoms and metal slabs, the Hagenmuller-Delmas' notation [35] is applied. Therein, the $Li_2Ni_2TeO_6$ lattices are classified as 'T2-type' structures where the letter 'T' denotes the tetrahedral structure and the number '2' indicates the number of layers per unit cell. Similarly, the Na octahedral and prismatic structures mentioned above are designated as 'O3-type' and 'P2-type' structures respectively. [1]

In general, 'O-type' and 'P-type' honeycomb layered oxides are considered to be superior cathode materials for high-performance energy storage systems, as their weak interlayer bonds readily create vacancies in the transition metal slabs that enable facile alkali-ion diffusion within the layers.[32] Besides the enhanced alkali-ion kinetics, fascinating structural disorders characterised by shears on transition metal slabs or shifts in stacking orders are induced during the electrochemical alkali extraction and reinsertion leading to a manifold of physicochemical properties such as an assortment of voltage-capacity profiles, improved rate performance and capacity retention (cyclability). [1] For instance, during battery operations, $Na_3Ni_2SbO_6$ and $Na_3Ni_2BiO_6$ cathodes have been observed to shift from the initial O3-type structure to a P3-type structure and eventually into an O1-type structure, resulting in phase transitions manifested by staircase-like voltage profiles. [22,27,33,37] As might be expected, due to the weaker atomic structures of P-type materials such as $Na_2M_2TeO_6$, structural disorders tend to be more prevalent, leading to better electrochemical performance as seen in $Na_2Ni_2TeO_6$, $Na_2Zn_2TeO_6$ and $Na_2Mg_2TeO_6$. [3-7,36-43]

As structural disorders can either be deleterious or beneficial to the functionality of layered materials, understanding their nature is pertinent not only as an avenue for fine-tuning emergent properties, but also as a means of unearthing new functional attributes such as magnetism and related microscopic phenomena. In layered materials, stacking disorders in the arrangement of the layers (stacking faults) occur predominantly perpendicular to the slabs, enlisting a variety of dislocations in some domains, particularly edge and screw dislocations. [57,58] Such dislocations are uniquely identified by two characteristic vectors, namely the Burgers vector and the sense vector (points along the dislocation line). For instance, in edge dislocations, these vectors are perpendicular to each other, thus requiring regions of shear, strain and stress to form in the crystal. On the other hand, in screw dislocations, these vectors are parallel to each other, thus precluding any regions of stress and strain from forming. The utility of using Burgers vectors to identify and characterise dislocations lies in the observation that Burgers vectors are conserved along the dislocation line, even during plastic deformations during such processes as (de)intercalation in cathode materials.

To explore structural changes in such cathode materials, X-ray diffraction (XRD) and neutron diffraction measurements are conducted on the specimen after undergoing electrochemical reactions in battery operations to show the structural changes occurring within the material. However, the limited resolution of these crystallographic analyses cannot account for the instantaneous structural evolutions occurring shortly after synthesis of the material. As a solution, transmission electron microscopy (TEM) can be employed alongside the aforementioned crystallographic analyses to provide local atomistic information that would not only show the existence of structural disorders but also provide information on the nature of the structural changes related to synthesis of the material.

In this context, theoretical and experimental studies to investigate the structural disorders particularly on the aforementioned O-type and P-type Na honeycomb layered oxides have been commissioned in an attempt to draw correlations between the atomic structure (microscopic details) with macroscopic manifestations such as electrochemical performance and phase transitions. [44-46] Recent studies utilising TEM on the O3-type $Na_3Ni_2SbO_6$ have revealed the existence of disordered sequences in the arrangement of Ni and Sb atoms as a possible reason behind phase transitions and improved ion mobility. [44] Even though, P-type $Na_2M_2TeO_6$ would make an ideal platform for investigating structural revolutions on pristine cathodes, crystallography studies on these materials lack rigour with literature limited only to theoretical computations and XRD analyses conducted on $Na_2M_2TeO_6$. [45-47] As such, information pertaining the local atomistic structures (TEM) and emergent attributes remain elusive.

Therefore, to investigate the nature of stacking disorders in $Na_2M_2TeO_6$ P-type structures, we utilise atomic-resolution scanning transmission electron microscopy (STEM) to illuminate the local structural disorders innate in pristine $Na_2Ni_2TeO_6$. We unveil a multitude of stacking faults of the metal slabs along the *c*-axis, revealing domains that manifest a variety of superstructures. We further discover variations in the distribution of the Na atoms in adjacent layers ascribed to aperiodic shifts along the layer stacking direction (*id est*, [001] zone axis). Finally, a supercell is proposed to demonstrate the arrangement of Na atoms within the *ab* plane.

## 2. RESULTS

### 2.1. CHARACTERISATION OF PRISTINE (AS-PREPARED) $Na_2Ni_2TeO_6$

The pristine $Na_2Ni_2TeO_6$ prepared via the high-temperature ceramics route, as described in the **Experimental** section, displays excellent crystallinity. The X-ray diffraction (XRD) patterns were explicitly indexed and refined to the hexagonal layered structure adopting the *P*6$_3$/*mcm* space group (**Figure 1a**). The refined lattice parameters ($a$ = 5.2049 (1) Å, $c$ = 11.1505 (5) Å) are in good accord with previously reported values.[3,6,36,47,50] Scanning electron microscopy, more specifically, energy-dispersive X-ray spectroscopy (EDX) was used to verify the stoichiometry and homogeneous elemental distribution of pristine $Na_2Ni_2TeO_6$ as shown in the supplementary information (**Figure S1, Figure S2** and **Table S1**). As illustrated in **Figure 1b,** the refined layered crystal

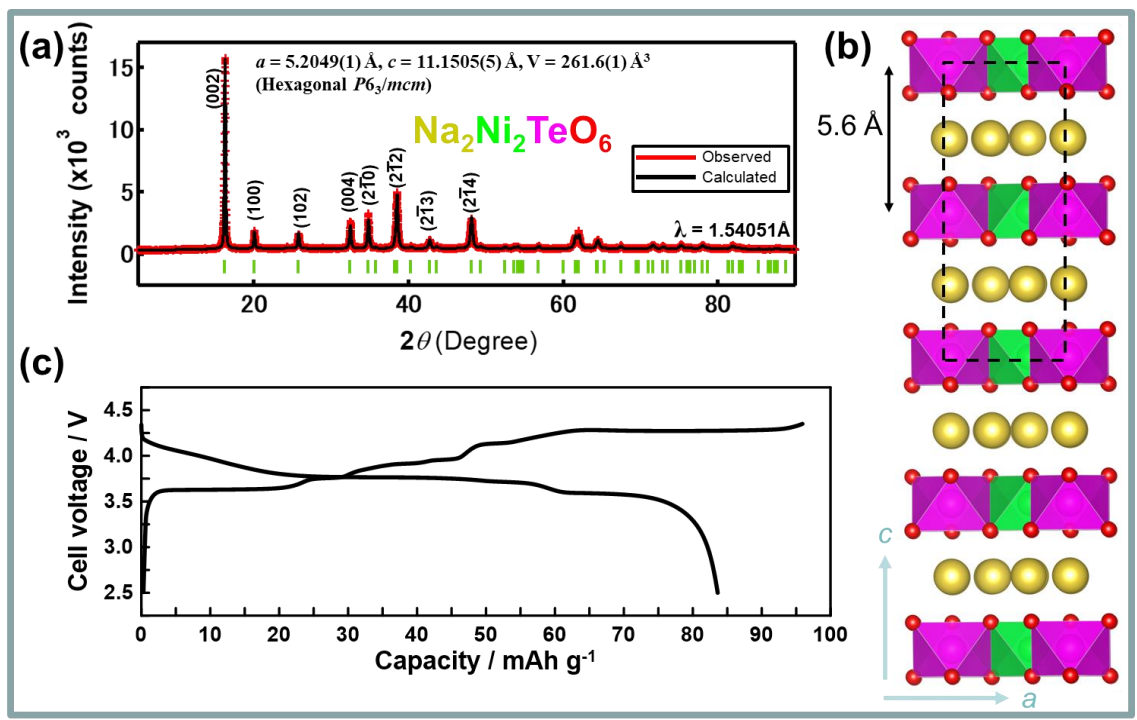

**Figure 1** Physicochemical characterisation of as-prepared $Na_2Ni_2TeO_6$ honeycomb layered oxide. **(a)** X-ray diffraction (XRD) pattern of as-prepared $Na_2Ni_2TeO_6$ indexed in the hexagonal *P*6$_3$/*mcm* space group. **(b)** A polyhedral representation of the crystal structure of $Na_2Ni_2TeO_6$ derived from the XRD refinement. Na atoms and oxygen atoms

structure of $Na_2Ni_2TeO_6$ entails Na atoms (in yellow) sandwiched between metal slabs comprising $NiO_6$ (green) octahedra surrounded by multiple $TeO_6$ (pink) octahedra. Each Ni and Te atoms are coordinated to six oxygen atoms in their respective octahedra whereas the Na atoms coordinate with oxygen atoms from adjacent metal slabs to form a prismatic structure with two repetitive Na atom layers per unit cell (defined as P2-type stacking), as shown in inset of **Figure 1b**. Galvanostatic (dis)charge measurements performed on pristine $Na_2Ni_2TeO_6$ (**Figure 1c**) show typical staircase-like voltage profiles, in concordance with voltage-capacity profiles reported for the P2-type framework.[36,50]

For an explicit analysis of the stacking sequences and honeycomb ordering, aberration-corrected scanning transmission electron microscopy (STEM) was performed on the synthesised $Na_2Ni_2TeO_6$ samples.[51-53] **Figure 2a** shows a high-angle annular dark-field (HAADF) STEM image of pristine $Na_2Ni_2TeO_6$ as viewed along the [100] zone axis. The contrast ($I$) of the HAADF-STEM images are proportional to the atomic number ($Z$) of elements along the atomic arrangement (where $I \propto Z^{1.7} \approx Z^2$) for clarity.

From the $b$-axis, the alignment of Te atoms represented by bright yellow spots ($Z = 52$) and Ni atoms marked by darker amber spots ($Z = 28$), manifest a Te–Ni–Ni–Te sequence (shown in **Figure 2a** inset) as should be expected from a P2-type honeycomb structure. The placement of atoms observed along the [100] zone axis is further validated by elemental mapping by STEM-EDX as shown in **Figure S3**. In addition, Na-atom layers interposed between the Ni and Te slabs can be discerned from the corresponding annular bright-field (ABF) STEM images shown in **Figure 2b**. As for ABF-STEM images, $I \propto Z^{1/3}$, which means that elements with lighter atomic mass such as Na ($Z = 11$) and O ($Z = 8$) can be visualised. A crystal model derived from the above-mentioned TEM measurements explicitly shows the P2-type framework of $Na_2Ni_2TeO_6$ as viewed along the [100] zone axis **(Figure 2c)**. It is worth noting that the varying intensity of the Na atom layers as seen in the ABF-STEM images (**Figure 2b**), evince the occupation of Na atoms in distinct crystallographic sites with varying occupancies; typical for this class of tellurates, as ascertained by the Rietveld refinement results shown in **Table S1.** In addition, there are different contrasts at Na sites that should be crystallographically equivalent (shown in red circles), indicating a modulation in the occupancies as shall be discussed in detail in a later section. To confirm the 3D structure model and the P2-type framework, STEM images of the same crystal were taken along the [1-10] zone axis as shown in **Figures 2d** and **2e.** The ABF-STEM image (**Figure 2e**) not only affirm the varying occupancy of Na atoms, but also highlight the zigzag arrangement of oxygen atoms in the adjacent metal slabs, a characteristic of the P2-type stacking, depicted in the crystal model (**Figure 2f**).

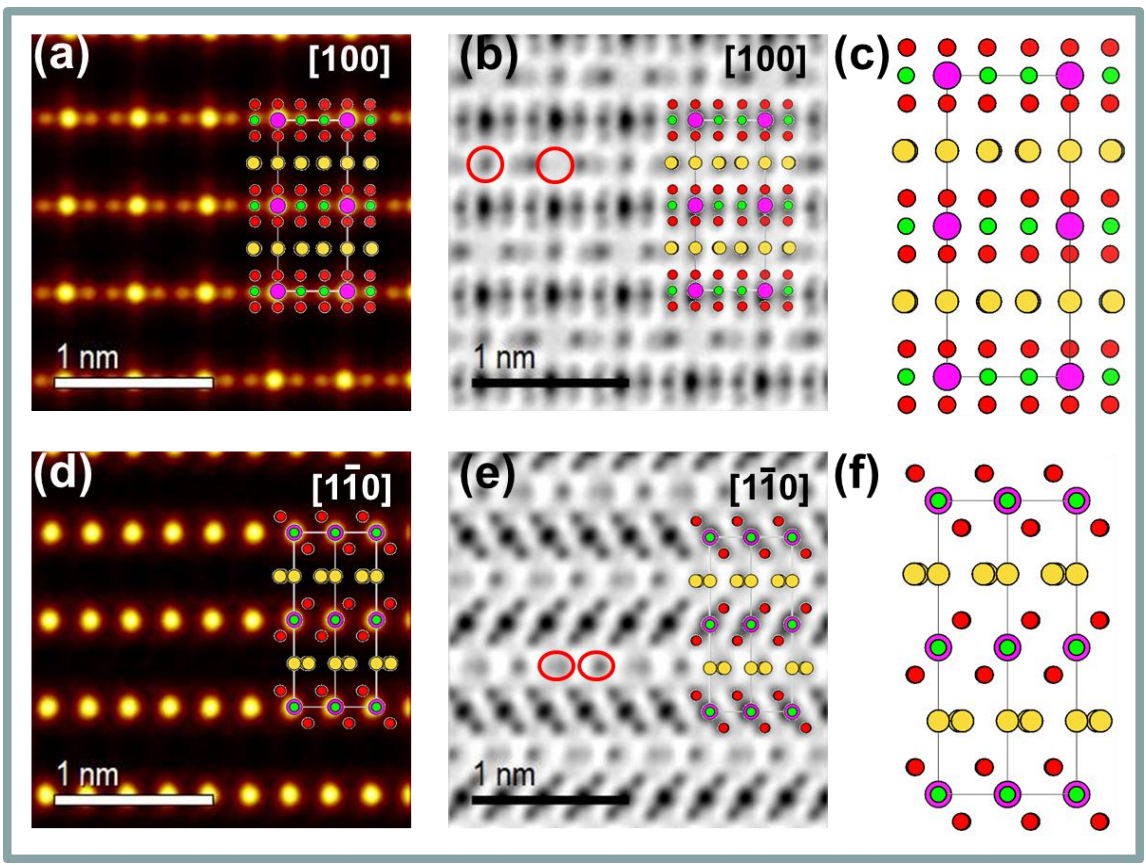

**Figure 2** Arrangement of the atoms along the [100] and [1$\bar{1}$0] zone axes in the P2-type $Na_2Ni_2TeO_6$. **(a)** High-angle annular dark-field scanning transmission electron microscopy (HAADF-STEM) image of $Na_2Ni_2TeO_6$ taken along [100] zone axis showing the ordering sequence of Ni and Te atoms corresponding to the P2-type stacking. Inset shows a projected model of the crystal structure, for clarity. **(b)** Annular bright-field (ABF)-STEM image taken along [100] zone axis displaying the arrangement of sodium atoms. Red open circles illustrate the varying contrast of Na sites that are supposed to be crystallographically homologous. **(c)** A rendering of the P2-type stacking of $Na_2Ni_2TeO_6$ along the [100] direction. **(d)** Visualisation (along the [1-10] zone axis) using HAADF-STEM, and **(e)** Corresponding ABF-STEM image. **(f)** Projection of the crystal structure along [1-10], affirming the projected model of P2-type stacking of atoms as shown in **d** and **e**.

## 2.2. STACKING FAULTS DUE TO SHEAR TRANSFORMATIONS OF THE METAL SLABS

For a closer look into the arrangement of the metal slabs, HAADF-STEM and ABF-STEM images were taken along the [100] zone axis as illustrated by **Figures 3a** and **3b.** Despite the ordered arrangement of metal slabs in the P2-type framework, disorders in the arrangement of the metal slabs were observed. The HAADF-STEM images demonstrate an ordered structure, whereby Te atoms (bright yellow spots) are positioned directly below or above the adjacent slabs in a perfect vertical array as can be seen in **Figure 3a**. However, the slabs are observed to deviate from the vertical arrays in certain domains (as highlighted by the green lines), indicating the occurrence of stacking faults across the slab stacking direction (*c*-axis) characterised by shear transformations. The density of the slab stacking faults was also found to vary between the specimen particles. It is worth noting that in the area containing high density stacking faults, local orderings spanning over a range of about 9 layers (5 nm) were discovered as shown in **Figure 3c**. Along the stacking faults, the presence of two types of superstructured domains (*viz*., type-1 and type-2) was discerned. As expressed by the Hägg symbol, the type-1 superstructure is denoted as +–+–+–+–+ (where + and – denotes the left and right shift of Ni/Te atoms in the adjacent slabs, respectively) whereas the type-2 superstructure is designated by +0+0+0+0+0. A multitude of such superstructures were observed, and further confirmed by the corresponding ABF-STEM images shown in **Figure 3d**. However, in the HAADF/ABF-STEM images taken along [1-10] at the same area of specimen, the slab stacking faults are invisible as displayed in **Figure S4**; an important indication of the slab shear transformations due to edge dislocations present in the specimen, as will be detailed in the **DISCUSSION** section.

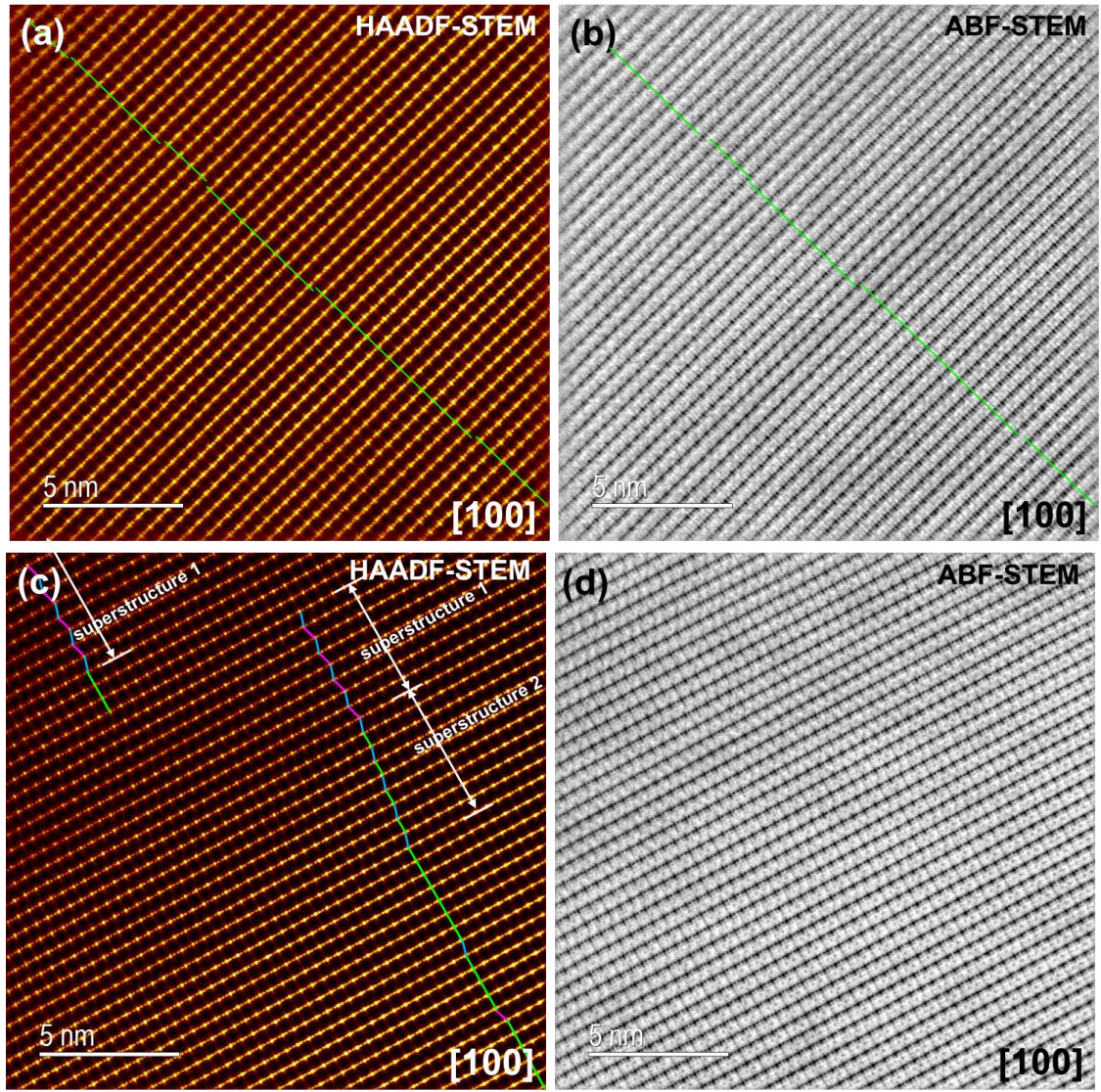

**Figure 3** Stacking disorders of metal slabs along the [100] zone axis in the P2-type $Na_2Ni_2TeO_6$ and their visualisation. **(a)** HAADF-STEM image of $Na_2Ni_2TeO_6$ taken along [100] zone axis showing faults in the stacking sequence of Ni and Te atoms. Green line serves as a guide for the readers. **(b)** Corresponding ABF-STEM image. **(c)** HAADF-STEM image of domains highlighting the presence of superstructures with alternating shift of the metals slabs along the *c*-axis ([001]). **(d)** Corresponding ABF-STEM image.

## 2.3. DISORDER IN THE STACKING OF SODIUM ATOMS

The occurrence of multiple disorders involving shifts in the metal slab layers along the *c*-axis not only reflects the diversity of the disorders innate in $Na_2Ni_2TeO_6$, but may also be envisioned to induce disorders in the arrangement of Na atoms sandwiched between the slabs. To investigate this hypothesis, STEM images taken along the [100] zone axis (**Figures 4a** and **4b**) were analysed. In the HAADF-STEM images shown in **Figures 4a**, the metal slabs are seen to lie directly below or above one another in a vertical array

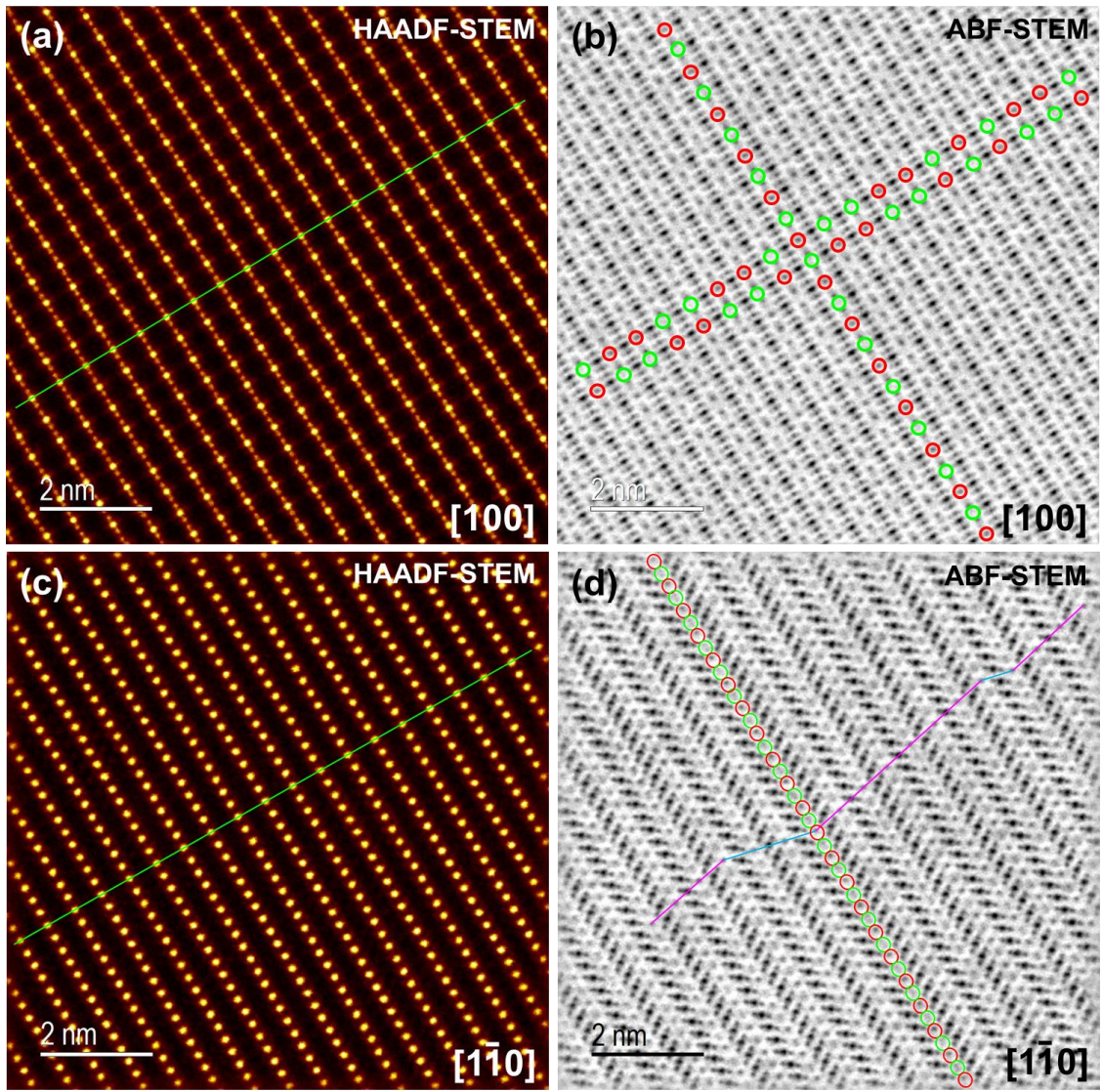

**Figure 4** Stacking arrangement (sequences) of Na atoms spotted along the [100] and [1$\bar{1}$0] zone axes in the P2-type $Na_2Ni_2TeO_6$. (**a**) HAADF-STEM image of $Na_2Ni_2TeO_6$ taken along [100] zone axis showing a perfectly ordered arrangement of metal slabs along the *c*-axis. (**b**) Corresponding ABF-STEM image showing a repetitive

sequence in the arrangement of Na atoms along the *b*-axis (with a periodicity of 2). The green and red circles show repetitive sequences in the Na sites judging from the similarity in the intensity contrast (thus occupancy). No periodicity is observed in the arrangement of Na in their respective sites along the *c*-axis. **(c)** HAADF-STEM image taken along the [1-10] showing a domain with a regular sequence in the arrangement of Te atoms along the *c*-axis and **(d)** Corresponding ABF-STEM images taken along the [1-10] zone axes showing double periodicity within the *ab* plane and stacking disorder along the *c*-axis of the sites occupied by Na (coloured guidelines link Na sites with similar contrast) which corresponds to the stacking disorder observed along the *c*-axis in (b).

across the slab (as further depicted by the green line in **Figure 4a**). However, in the ABF-STEM images (**Figure 4b**), where lighter atomic mass elements such as Na can also be distinguished, a modulation of Na occupancy is observed as illustrated by the red and green circles which represent the different crystallographic sites previously established (**Figure 2b**). Along the directions in the *ab* plane (perpendicular to the *c* axis), the red and green circled arrays are seen to vary in contrast with a high regularity of alternation, indicating the presence of a superlattice with the double periodicity. However, there is no coherency in the modulations of Na atoms along the *c*-axis, as the phase of the modulation between adjacent Na planes is observed to frequently invert with no periodicity. For better understanding of the Na layer disorders, additional STEM images taken along the [1-10] zone axis corresponding to the crystallite viewed in a tilted angle of $30^0$, are shown in **Figures 4c** and **4d**. The ABF-STEM images (**Figure 4d**) clearly showcase the highly ordered double periodicity in the *ab* plane and the aperiodicity along *c*-axis (clearly mapped out by the coloured lines linking the Na sites with the same contrast).

## 2.4. ELECTRON DIFFRACTION PATTERNS TAKEN ALONG MULTIPLE ZONE AXES IN $Na_2Ni_2TeO_6$

To validate the results of STEM, electron diffraction measurements were performed along [100] and [1-10] zone axes. To gain insight into the selected area electron diffraction (SAED) results, it is important to reiterate that $Na_2Ni_2TeO_6$ crystallises in a hexagonal lattice with cell dimensions shown in **Figure 1a**. The SAED patterns taken in the area without slab stacking fault along the [100] and [1-10] zone axes (**Figures 5a** and **5d**) exhibit clear diffraction spots that are indexable to the hexagonal cell as shown in the kinematically simulated pattern (**Figures 5b** and **5e**). The appearance of a 'streak-like' array of spots (shown in green) instead of discrete spots confirms the stacking disorder (faults) of Na planes along the *c*-axis. However, 'streak-like' patterns could not be reproduced by the kinematically simulated patterns, indicating the existence of super-periodicity along the *ab* plane. Thus, a 2*a* supercell model illustrated by **Figures 5c** and **5f** was considered. The model reproduces the arrays of spots at positions identical to the experimental electron diffractograms confirming the double super-periodicity of Na occupancy. It is worthy to note that intensity distribution of the streaks in the experimental diffractograms is completely different from those in the simulation models. This is attributed to the fact that our model does not consider the stacking faults of Na superlattice planes along the *c*-axis.

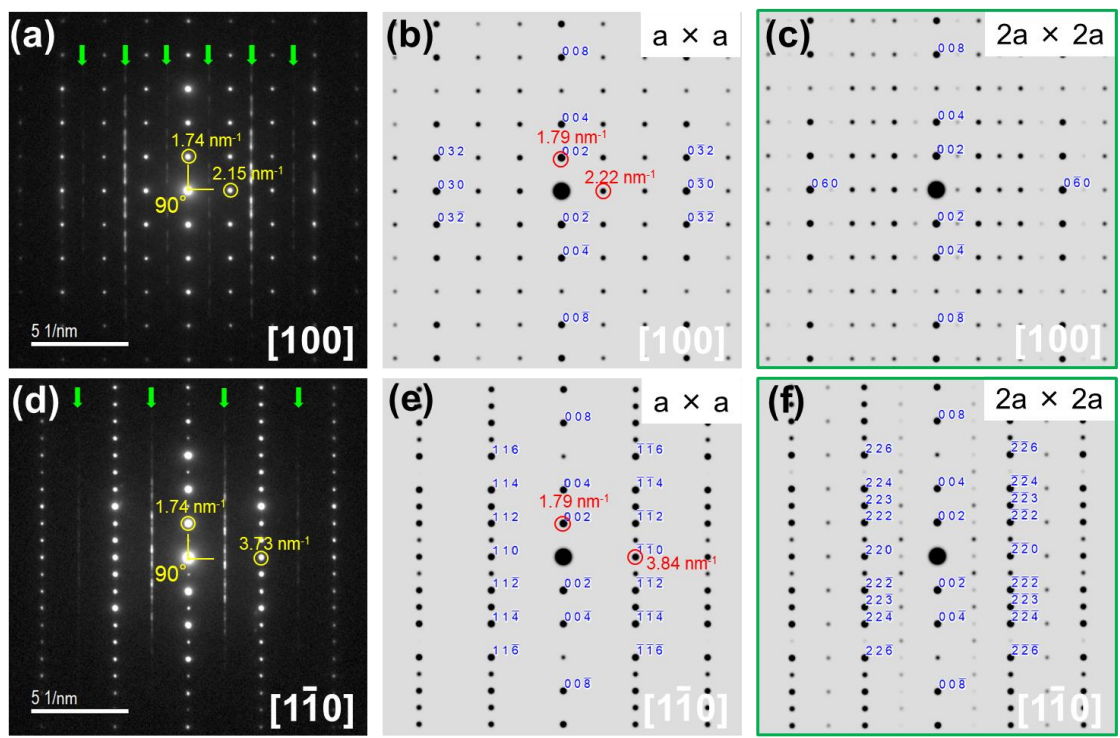

**Figure 5** Comparison of the electron diffractograms of $Na_2Ni_2TeO_6$ based on the original cell and the proposed supercell. **(a)** Selected area electron diffraction (SAED) patterns of $Na_2Ni_2TeO_6$ taken along the [100] zone axis highlighting streaks (green arrows) in the diffractograms hallmarking the presence of a supercell. **(b)** Simulated diffractograms along the same [100] axis using the original cell and **(c)** Supercell with a manifold dimensions of the unit cell along the *a*-axis and *b*-axis, which reproduces the experimentally obtained electron diffraction patterns. **(d)** SAED patterns of $Na_2Ni_2TeO_6$ taken along the [1-10] also underpinning streaks in the diffractograms. **(e)** Simulated diffractograms along the same [100] axis using the original cell and **(f)** Supercell with a manifold dimensions of the unit cell along the *a*-axis and *b*-axis, which reproduces the experimentally obtained electron diffraction patterns.

## 3. **DISCUSSION**

$Na_2Ni_2TeO_6$ crystallises in an ordered P2-type layered framework with Te and Ni atoms positioned vertically above and below in the adjacent slabs. High-resolution STEM reveals disorders engendered by shifts across the honeycomb slabs. The lack of periodicity across the slab (*i.e.*, along the *c*-axis) is attributed to disorder in the position of metal atoms (Ni and Te) that is exacerbated by the inherently weak interlayer bonding between Na atoms and the adjacent metal slabs. Slab stacking faults were seen to infiltrate the entire crystallite specimen, with the exception of some localised (closed) ones which were confined in domains illustrated by the gap between the green lines in the HAADF-STEM images taken along the [100] zone shown in **Figure 6a**. A better understanding of the nature of the slab stacking faults was obtained when the crystallite was tilted by $30^0$ and thereafter, STEM images taken along the [1-10] zone axis as shown in **Figures 6b** (The corresponding ABF-STEM images have been furnished in **Figure S5**). The locality of these gaps is expanded in **Figure 6c** and **Figure 6d** to reveal the shear transformations indicated by the blue arrows, and the position of the edge dislocation marked by a 'T' corresponding to Burgers vectors, $p_1$ and $p_2$ viewed from the $[100]$ and [1-10] zone axes respectively. Such edge dislocations correspond to the insertion of an additional lattice plane along *c* axis at the end of the stacking fault.

As such, for any arbitrarily chosen Te/Ni metal slab where shear transformations are observed, the layers move one metal (*i.e.*, Ni (in green)) to a position where the other (*i.e.*, Te) normally would be expected to be situated in the relative slab. Consequently, the Burgers vectors, $p_1$ and $p_2$ can be determined to be [1/3 –1/3 0] and [2/3 1/3 0] respectively (**Figure 6e** and **Figure 6f**). It is worthy to note that the resulting superstructured domains entailing the shear transformations could be indicative of a symmetry change and is thus a subject of future theoretical and experimental studies. SAED patterns of the area containing the superstructured domains (furnished in **Figure S6**) reveal diffraction spots that are not permitted in the $P6_3/mcm$ (centrosymmetric) space group originally indexed for the pristine $Na_2Ni_2TeO_6$. The (0*kl* with $l = 2n +1$) diffraction spots seen in the SAED patterns are allowed in hexagonal space groups such as $P6_322$.

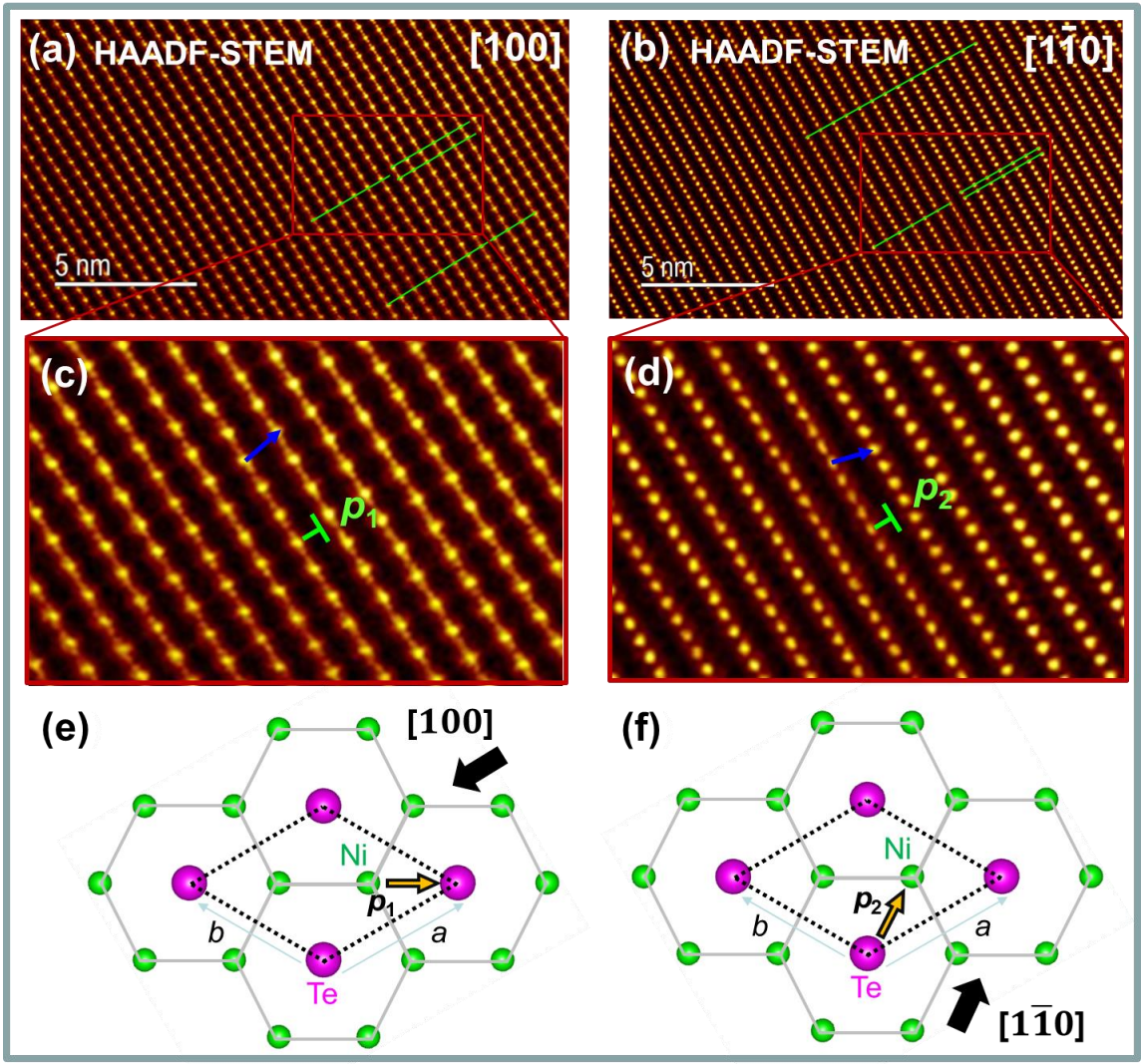

**Figure 6** Stacking fault of the metal slabs terminated by two partial dislocations observed along the [100] and [1$\bar{1}$0] zone axes in the P2-type $Na_2Ni_2TeO_6$. **(a)** HAADF-STEM image of $Na_2Ni_2TeO_6$ taken along [100] zone axis showing the location of the edge dislocation in the arrangement of Te and Ni atoms relative to their corresponding positions in the adjacent slabs along the *c*-axis. Green lines serve as guide for the eye where the broken lines indicate the presence of a shear transformation. **(b)** HAADF-STEM image of $Na_2Ni_2TeO_6$ taken along the [1-10] zone axis showing the location of the edge dislocation in the arrangement of Te and Ni atoms relative to their corresponding positions in the adjacent slabs along the *c*-axis. To reiterate, green lines serve as guide for the eye where the broken lines indicate the presence of a shear transformation. **(c)** and **(d)** The locality of the edge dislocations expanded to reveal the shear transformations indicated by the blue arrow, and the position of the dislocation marked by a 'T' corresponding to the Burgers vectors, $p_1$ and $p_2$ as viewed from the [100] and [1-10] zone axes respectively. **(e)** and **(f)** Schematic illustrations of the corresponding edge dislocations in the arrangement of Te and Ni atoms relative to their positions in the adjacent slabs along the *c*-axis as observed along the [100] and [1-10] zone axes (black arrows) respectively. The Burgers vectors, $p_1$ and $p_2$, can be determined to be [1/3 –1/3 0] and [2/3 1/3 0] respectively. For clarity, Te atoms are shown in pink whilst Ni atoms are in green.

Typical disorders observed in honeycomb layered oxides involve shifts of the metal slabs, as observed in materials such as the O3-type $Na_3Ni_2DO_6$ (*D* = Bi and Sb)[33,54] as well as twinning of the metal slabs have been found amongst materials such as $Cu_3M_2SbO_6$ (*M* = Ni and Co) and their derivatives.[55,56] A few reports on disorders embodied by skipping or disappearance of the stack layers (or what is referred to as dislocations) in oxide materials have also been recently availed.[57,58] Thus, the edge dislocations observed in $Na_2Ni_2TeO_6$, can be seen as an indication of the existence of new topological disorders such as curvature which may rationalise the superior physicochemical properties of $Na_2Ni_2TeO_6$.

Moreover, the disordered arrangements seen in the Na occupancy sites can significantly alter the $Na^+$ kinetics and ionic conductivity displayed by $Na_2Ni_2TeO_6$. Thus, related honeycomb layered oxide materials that display high ionic conductivity and superb kinetics can be subject of future study to ascertain the existence of any correlations to the observed stacking faults and dislocations reported herein. With the aid of electron diffraction, an appropriate superstructural model was proposed to determine the periodicity of the modulation for Na atoms in $Na_2Ni_2TeO_6$ (**Figure 7**). As expected, Na atoms are distributed in three crystallographic sites with varying occupancies (as illustrated in **Table S1**).

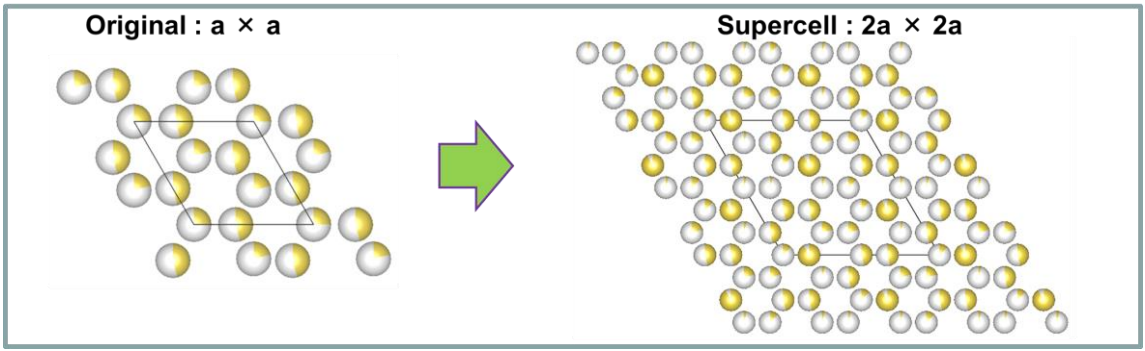

**Figure 7** Arrangement of the Na site occupancy within an *ab* plane in $Na_2Ni_2TeO_6$ based on the original cell derived from the XRD and the proposed $2a \times 2a$ supercell.

Examination of the Na sites along the *ab* plane in the original P2-type framework (**Figures 2, 4b** and **4d**), reveals a highly ordered double periodicity, apparent in both the [100] and $[1\bar{1}0]$ axes. This suggests that a $2a \times 2a$ supercell represents the most appropriate dimensions to describe the arrangement of Na atoms in $Na_2Ni_2TeO_6$ along the *ab* plane. This model of Na occupancy in the *ab* plane can be used to rationalise the contrast variations seen in the ABF-STEM in both crystal orientations. The $2a \times 2a$ model is further validated by the simulated kinematical diffraction pattern highlighted in **Figure 5**. ABF-STEM images show that the Na superstructure in the *ab* plane and its stacking faults are always present with or without stacking fault of the slab, which can also be confirmed by electron diffraction of domains with dense slab disorders (**Figure S6a** and **S6b**) as well as those without any slab disorders (**Figures 5a** and **5d**). $Na_2Ni_2TeO_6$ is characterised by aperiodicity due to multiple stacking faults along the *c*-axis, as opposed to the complex but highly ordered 2D structure in the *ab* plane. Nonetheless, regardless of the precise nature of the model required to accurately capture the Na occupancies, the existence of such a large ordered two-dimensional (2D) structure in the *ab* plane represents an intriguing result which necessitates further inquiry into the precise role played by the stacking disorders reported herein.

# 4. CONCLUSION

In conclusion, this study clearly demonstrates the efficacy of atomic-resolution transmission electron microscopy (TEM) in unravelling the defect structures of related honeycomb layered oxides, in particular the pristine $Na_2Ni_2TeO_6$. Detailed TEM analyses provide a novel outlook of the local atomistic structures, revealing the coexistence of stacking faults of metal Ni/Te slabs described by shifting of the slab layers, alongside the $2a \times 2a$ superstructure of Na site occupancy in *ab* plane and its stacking fault along the *c*-axis. The existence of superstructured domains in $Na_2Ni_2TeO_6$ not only opens up avenues for fascinating research into the structural disorders inherent in pristine tellurates, but also establishes $Na_2Ni_2TeO_6$ as a model honeycomb layered oxide material to study innumerable defects with possible implications for their functionality as cathode materials.

## Declaration of Competing Interests
The authors declare no competing interests.

## Acknowledgements
T.M. thanks Ms. Shinobu Wada and Mr. Hiroshi Kimura for the unrelenting support in undertaking the entire study. T. M. also gratefully acknowledges Ms. Kumi Shiokawa, Mr. Masahiro Hirata and Ms. Machiko Kakiuchi for their advice and technical help as we conducted the syntheses, electrochemical and XRD measurements. This work was conducted in part under the auspices of the Japan Society for the Promotion of Science (JSPS KAKENHI Grant Number 19K15685), Sumika Chemical Analyses Services (SCAS) Co. Ltd., National Institute of Advanced Industrial Science and Technology (AIST) and Japan Prize Foundation.

## Supplementary material
The online link to Supplementary material (which includes the experimental details) associated with this article can be accessed via the link:
https://www.sciencedirect.com/science/article/pii/S2589152921000065